# Bandwidth broadening of X-ray free electron laser pulses with the natural gradient of planar undulator

Minghao Song[a,b], Kai Li[a,b], Jiawei Yan[a,b], Han Zhang[a], Chao Feng[a], Haixiao Deng[a,*]

[a] Shanghai Institute of Applied Physics, Chinese Academy of Sciences, Shanghai, 201800, P. R. China

[b] University of Chinese Academy of Sciences, Beijing 100049, P. R. China

**Abstract**    Besides the target to pursue the narrow bandwidth X-ray pulses, the large bandwidth free-electron laser pulses are also strongly demanded to satisfy a wide range of scientific user experiments. In this paper, using the transversely tilt beam enabled by deflecting cavity and/or corrugated structure, the potential of large bandwidth X-ray free-electron lasers generation with the natural gradient of the planar undulator are discussed. Simulations confirm the theoretical prediction, and X-ray free-electron laser bandwidth indicates an increase of one order of magnitude with the optimized parameters.

## 1. Introduction

X-ray free-electron laser (FEL) facility as the primary candidate for fourth-generation light source that holds the promise to provide extremely bright X-ray photons with femtosecond pulse duration. The leading-edge instruments offer several capabilities in a wide range of scientific applications in chemistry, biology, material science and physics (Barletta *et al*., 2010; Öström *et al*., 2015). Currently, the operated or constructed FEL facilities around the world are generally based on the self-amplified spontaneous emission (SASE) (Emma *et al*., 2010; Ishikawa *et al*., 2012; Ackermann *et al*., 2007; Kondratenko & Saldin, 1980; Bonifacio *et al*., 1984) and seeded (Deng & Feng, 2013; Yu, 1991; Stupakov, 2009) FEL process. SASE FELs produce coherent radiation with final natural bandwidth between $10^{-3}$ and $10^{-4}$ (Bonifacio *et al*., 1984), while the seeded FELs even further reduce the natural bandwidth to the Fourier transform limit (Zhao *et al*., 2012; Amann *et al*., 2012; Saldin *et al*., 2001).

However, besides aiming at minimizing the bandwidth of the produced radiation, large bandwidth FEL pulse is also of great interest. With the help of the ultra-brightness of X-ray FEL and generation of broad bandwidth, it shows promise for revealing essence of molecular structures and solving remained problems, which will facilitate the progress in a variety of research fields, for example, studying microcrystalline materials (Catherine *et al*., 2015; Baradaran *et al*., 2013), analyzing single-cell (Patterson *et al*., 2010), determining multi-wavelength anomalous diffraction (MAD) phase information (Son *et al*., 2011). Moreover, compared with standard narrow bandwidth operation, large bandwidth FEL allows for wavelength tuning in a more flexible way. More recently, large bandwidth FEL schemes were newly proposed (Prat *et al*., 2016; Hernandez *et al*., 2016), which mainly relies on the electron beam energy chirp formed in linear accelerators and/or the transverse gradient undulator.

The generation of FEL relies on the relativistic electron beam moves along a periodic array of dipole magnets and wiggled transversely to emit electromagnetic radiation (Madey, J. M. J., 1971; Kondratenko & Saldin, 1980; Huang & Kim, 2007). According to FEL resonant condition, both time-energy correlation in the electron bunch and space-field correlations in the undulator can broaden the FEL bandwidth. Generally, the simplest and natural way to obtain large bandwidth FEL is using an energy chirped electron beam (Andonian *et al*., 2005). However, large energy chirp will be hindered and challenged for the X-ray FEL. Apart from it, there are many other approaches to produce the energy chirp, for instance, utilizing the space charge effects of an extremely compressed beam (Serkez *et al*., 2013), modifying the longitudinal laser profile of photo injector (Penco *et al*., 2014) and using the over-compressed beam combined with wakefields (Hernandez *et al*., 2016; Emma, 2000; Turner *et al*., 2015). Alternatively, the application of transverse gradient undulator may contribute a full bandwidth of 10% in soft X-ray of 1 nm wavelength (Prat *et al*., 2016; Bernhard *et al*., 2016). Motivated by these early works, in this paper, X-ray FEL bandwidth broadening enabled by the natural gradient (vertical gradient usually) in a planar undulator is studied utilizing currently existed instruments and mature accelerator technologies. It is demonstrated that a moderate RMS bandwidth above 0.4% could be obtained, which lead to an eight-fold enhancement in hard X-ray region of 0.1 nm. Moreover, if the X-ray pulse energy requirement is not very high in the user experiment, i.e., FEL saturation is not mandatory, RMS bandwidth as large as 2.0% can be achieved.

## 2. Principle of scheme

In this paper, in condition to the techniques mentioned above, a simple scheme which consists of transverse-longitudinal coupling section and planar undulator section is proposed. As illustrated in Fig. 1, a de-chirper device (Fu *et al*., 2015; Deng *et al*., 2014; Bane *et al*., 2016; Zagorodnov *et al*., 2015; Zhang *et al*., 2015; Emma *et al*., 2014; Novokhatski, 2015) or transverse deflecting structure (TDS) (Akre *et al*., 2001; Tan *et al*., 2014) is installed before the undulator to introduce a vertical tilt to the electron beam. It is expected that such a beam passes through a planar undulator and produce FEL radiation with large bandwidth. The initial electron beam has the following properties depending on the FEL lasing requirements, beam energy of 6 GeV and peak current of 3 kA can be delivered with 200 pC bunch charge, with the normalized emittance less than 0.3μm-rad to satisfy the desired photon output. In this work, it is convenient to change the gap of de-chirper or deflecting voltage of TDS, in this sense, the beam will experience different transverse kick and can be characterized with various tilt amplitudes, and thus one can simply tune the bandwidth of the final radiation. In principle, the corrugated structure could be used to further increase the beam energy chirp with the longitudinal wakefield, and thus the radiation bandwidth.

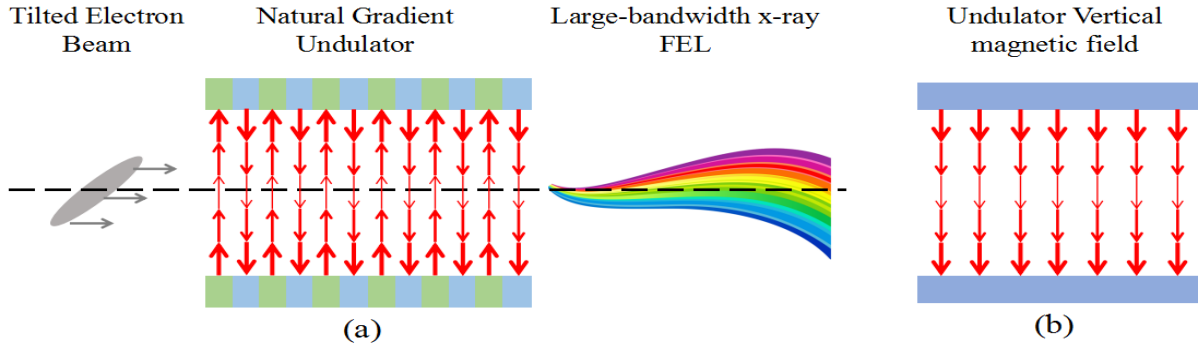

**Figure 1** The schematic layout of the proposed scheme: (a) the titled electron beam travelling through the planar undulator with natural gradient to produce broad bandwidth FEL pulse and (b) displays the vertical magnetic field of undulator cross section.

### 2.1. Generation of beam transverse tilt

As aforementioned, an electron beam with a transverse tilt travelling through a planar undulator produces broad bandwidth X-ray FEL pulse. The beam transverse tilt can be generated with several methods. For example, applying a RF deflector to kick beam or introducing the transverse wakefields of accelerating structures or corrugated structures in the FEL facility.

The RF deflecting cavities have been widely used in the beam diagnostics of FEL and many other accelerators (Korepanov *et al*., 2006; Park *et al*., 2015; Akre *et al*., 2001; Xiang *et al*., 2010; Ding *et al*., 2014; Lee *et al*., 2014). In the deflecting cavity, the deflecting voltage is zero for the longitudinal centre of the bunch and provides a linear transverse deflection to the rest of the bunch, according to the (Alesini *et al*., 2006), the transverse displacement of the bunch slice is proportion to the deflecting voltage, thus, the beam transverse tilt can be easily tuned through the deflecting voltage of the TDS structures. For example, a total power of 40 MW will be fed to a pair of X-band deflecting structures (Tan *et al*., 2014) for SXFEL. Therefore, a wide range of transverse tilt could be achieved. Here the Gaussian bunch distribution is typically taken into consideration to explain. Simulated images on the observation screen are presented in Fig. 2. It represents the beam image with vertical tilt amplitude of 0.5 mm. It can be achieved by a 40 MV deflecting voltage together with experiencing a 14 meters drift section, and verified by the ELEGANT (Borland, 2000) simulation.

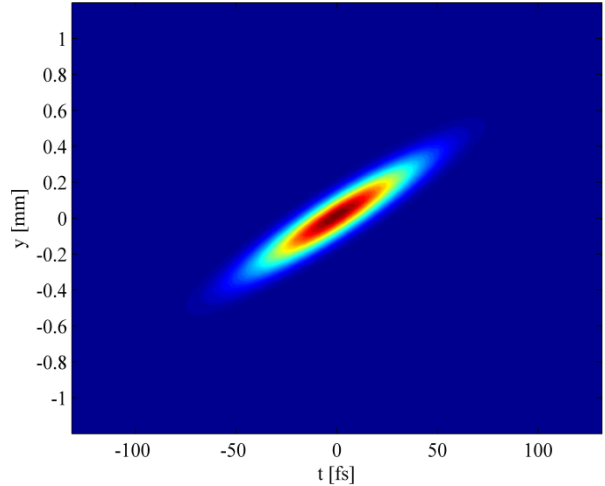

**Figure 2** The simulated image of Gaussian profile electron beam after the deflecting and drifting, the plot is the bunch phase space with vertical tilt amplitude of 0.5 mm. The bunch head is in the left.

A corrugated structure is usually a pipe or a pair of parallel plates with small bumps on the wall in a form of periodic ridges (Fu *et al*., 2015; Deng *et al*., 2014; Bane *et al*., 2016; Zagorodnov *et al*., 2015; Zhang *et al*., 2015; Emma *et al*., 2014; Novokhatski, 2015). For the large bandwidth FEL application the corrugated structure has small gap of millimetre order. When interacting with the longitudinal wakefields, the corrugated structures mainly responsible for beam energy chirp. While if the electron beam enters the structure with an offset from the axis, travelling close to one jaw of the structure, it excite the transverse dipole wakefields, where the bunch head receive no kick while the bunch tail will be transversely kicked. On the basis of the wakefield theory (Bane *et al*., 2016; Zhang *et al*., 2015; Novokhatski, 2015), it is interesting note that the bunch tail would experience a larger transverse kick with farther distance away from axis, in condition, the transverse tilt amplitude could be also simply tuned by varying the gap of the structure. Recently, on the basis of transverse beam tilt generated by the de-chirper at LCLS, a set of multi-colour X-ray FEL experiments was successfully accomplished (Lutman *et al*., 2016). In the broad bandwidth X-ray FEL case here, if a beam vertical tilt of ±0.5 mm is required, roughly it can be achieved when the electron beam passes through a 2.5-meter long LCLS-similar corrugated structure with a vertical offset of 0.64 mm.

### 2.2. Natural gradient of a planar undulator

Considering a planar undulator for which the magnetic field in the vertical direction:

$$B(y) = B_0 (1 + \frac{1}{2} k^2 y^2) \tag{1}$$

Here $k = 2\pi/\lambda_u$, $\lambda_u$ is the undulator period and $B_0$ is the peak magnetic field. When a relativistic electron entering into the undulator, it will wiggle periodically in the horizontal *x* direction and emit radiation with the resonant wavelength (Huang & Kim, 2007). If one neglects the high order term $k^4y^4$, one can derive that:

$$\lambda(y) = \lambda_0 + \frac{\lambda_u}{2\gamma_0^2} \cdot \frac{K_0^2}{2} \cdot k^2 y^2 \tag{2}$$

Here $K_0$ is the dimensionless undulator strength parameter and $\gamma_0$ is the electron energy in units of the electron rest energy. The FEL wavelength deviation in terms of the resonant wavelength $\lambda_0$ can be concluded:

$$\frac{\lambda(y) - \lambda_0}{\lambda_0} = \frac{K_0^2}{2 + K_0^2} k^2 y^2 \tag{3}$$

**2.3. Weak focusing lattice**

The electron beam size in the undulator has a great influence on the FEL gain length, on the one hand, too large beta function will enlarge the electron beam transverse cross section, which leads to the reduction of the current density and FEL gain, on the other hand, too small beta function will cause radiation diffraction and increase the gain length (Xie, 1995; Jia *et al.*, 2004). Therefore, optimization of the beam Twiss function is important for the final FEL performance. In general, the undulator section will be interrupted by several pairs of focusing quadrupoles and de-focusing quadrupoles to control the beam transverse size, which is well−known FODO lattice (Wiedemann, 2008). Commonly, according to the Eq. (3), it is not difficult to find that introduced spectral bandwidth is proportional to the square of *y*, which means the vertical distance away from the axis. Obviously, the strong vertical oscillation of beam motion has an impact on the exponential gain and coherence building, and further affects the generated FEL bandwidth. Therefore, a weak focusing FODO lattice is preferred here.

Fortunately, since the intrinsic vertical focusing effect of the planar undulator is quiet weak, a small vertical beta oscillation can be ensured and achieved. In the meantime, all quadrupoles in the undulator system are weak and horizontally focusing one to obtain a periodic and controllable beta vibration in the both horizontal and vertical direction. As shown in Fig. 3, an equivalent FODO lattice with length of 10 m consisting of focusing quadrupole, drift and a pair of planar undulator is considered, of which the undulator is 4.5 m long with period of 15 mm and $K_0$ parameter of 1.47. Based on the transform matrix of FODO cell, the properties of the beta function can be derived. Under such FODO cell, it is deserved to note that $\beta_y$ variation is well controlled according to Fig. 3. More specifically, the difference between the maximum and minimum $\beta_y$ is not beyond the 2 m, which is truly small compared with the averaged $\beta_y$ about 230.

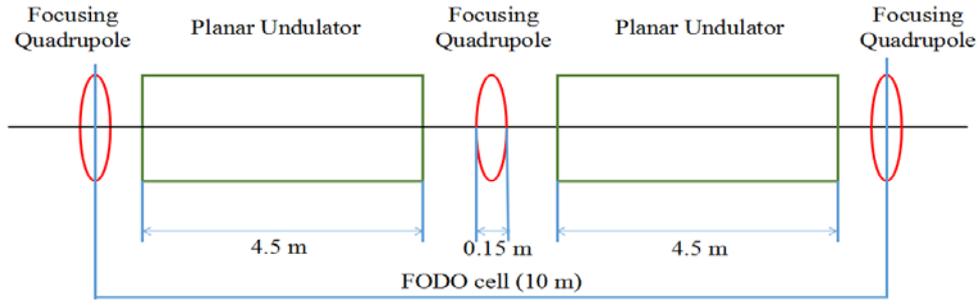

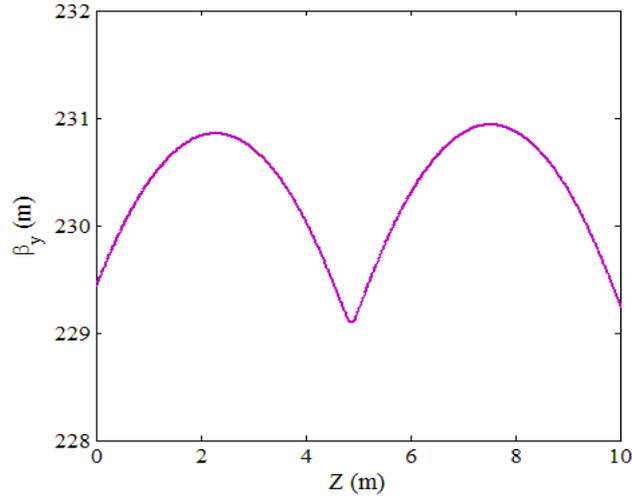

**Figure 3** The equivalent FODO cell and corresponded vertical beta oscillation along the longitudinal direction.

**2.4. Control of beam tilt amplitude**

For the proposed scheme in Fig.1, according to the aforementioned, the beam vertical oscillation is indeed not so strong because of a weak focused lattice. However, due to the residual vertical focusing strength, vertical kick given by the TDS system and/or transverse wakefields, the electron beam will still slowly varies its vertical size in the planar undulator, hence the generated FEL bandwidth would be narrowed and deviated from the expected. Fortunately, one can minimize the beam tilt variation by controlling the parameters of the whole system.

The undulator system can be roughly treated as an equivalent quadrupole with the focusing strength $k_f$ which is determined by the detailed FODO structure, and the total length of the undulator system $L$. If we consider an electron beam slice with coordinates of ($y_0$, $y_0'$, $z_0$) at the undulator entrance, it will experience a continuous vertical kick because of non-zero $y_0$, and the vertical motion can be stated as:

$$y'' = -\frac{ek_f}{cm}y \tag{4}$$

Where *e*, *m* are the charge and mass of an electron and *c* is the speed of light. Considering that the vertical motion of this slice is under well control in the whole undulator, i.e., the vertical kick given by undulator lattice is supposed to be constant, and then the vertical motion can be expressed as:

$$y(L) = -\frac{ek_f}{2cm}y_0 L^2 + y_0' L + y_0. \tag{5}$$

Now in Eq. (5), the first term represents the beam vertical tilt variation induced by the undulator lattice, while the second and the third terms together shows the beam tilt previously produced by the transverse deflector. In general, one can optimize the undulator lattice and transverse deflector parameters to minimize the beam tilt variation in the whole undulator. In our case, the beam tilt amplitude variation can be controlled within 2%.

### 3. Simulation results

In this work, the dimensionless undulator parameter $K_0$ is 1.47 and the undulator period is 15 mm, which leads an FEL resonant wavelength of 0.1 nm. When substituting these parameters into Eq. (3), one can derive the RMS FEL bandwidth

$$\frac{\Delta\lambda}{\lambda_0}[\%] \approx 2.73 y^2 [mm^2] \tag{6}$$

The theoretical expression in Eq. (6) indicates that the produced FEL bandwidth is proportional to the square of vertical offset. Now it seems that a beam with different tilt amplitude travelling through the planar undulator will produce radiation with different bandwidth. Besides, numerical simulations also have been performed with the help of code GENESIS (Reiche, 1999) to further investigate the cases and compare with the theoretical studies, in which a flat-top beam current is assumed.

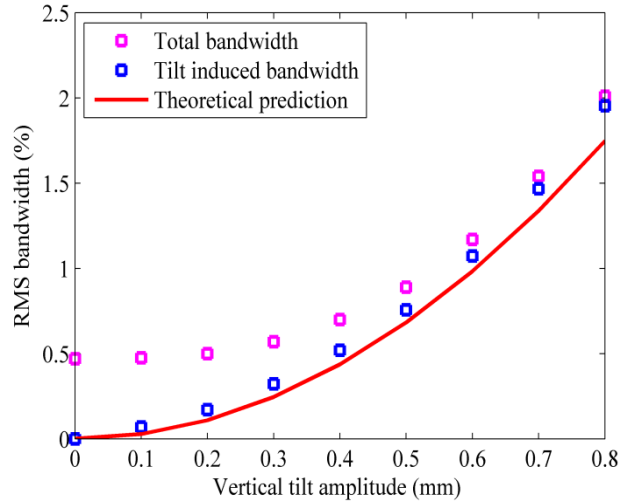

**Figure 4** RMS bandwidth after 10 m undulator versus the beam vertical tilt amplitude.

Time-dependent FEL simulations using different beam tilt amplitude are carried out to emulate the produced bandwidth. Figure 4 shows the RMS bandwidth after 10 m undulator where FEL just turns into the exponential gain region, far from saturation. The purple squares show the gross bandwidth from simulation, and there is still 0.5% bandwidth even if the electron beam is not vertically tilted, which is mainly contributed by the spontaneous emission. After the subtraction of this part, the net bandwidth induced by the vertical beam tilt can be obtained, as the blue squares shown in Fig. 4. In accordance with Eq. (6), it is not difficult to find that the net bandwidth is nonlinear to the tilt amplitude, in other words, is proportional to the square of tilt amplitude. Under this circumstance, the RMS output bandwidth may achieve 2.0% with a short undulator when an electron beam with 0.8 mm vertical tilt amplitude is used.

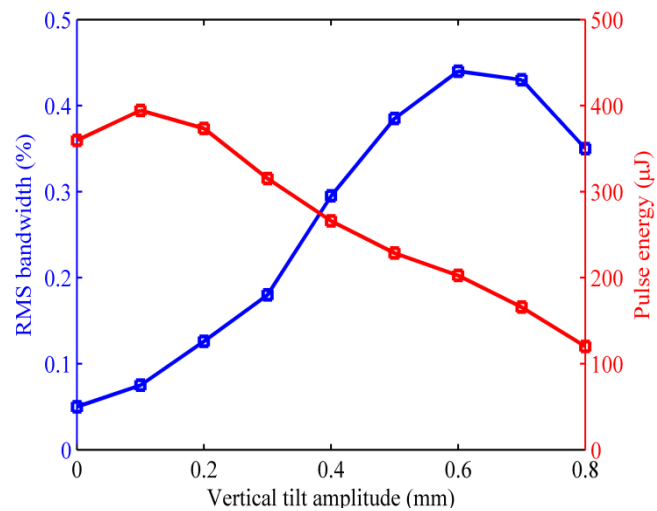

**Figure 5** RMS bandwidth and FEL pulse energy of 0.1 nm SASE at saturation.

A typical hard X-ray FEL undulator length is 100 m. After such an undulator, as illustrated in Fig.5, FEL pulse energy and accompanied RMS bandwidth as a function of the vertical beam tilt amplitude has been simulated. Generally, the behaviour of FEL pulse energy is reversed to the evolution of FEL bandwidth. For the pulse energy, it achieves a maximum of 395 μJ when the beam tilt amplitude is 0.1 mm. In this case, the ±0.1 mm beam tilt occasionally serves as an undulator taper for some appropriate part of the electron bunch. Then the pulse energy reduces approximately linearly with the increasing beam tilt amplitude, which can be easily understood. On the other hand, regarding to the bandwidth, it grows steadily from the initial 0.05%, and in particular reaches to maximum bandwidth of 0.44% at ±0.6 mm beam tilt. It is believed that, the electron beam is too much stretched to support a fruitful colour FEL lasing after ±0.6 mm vertical tilt and the bandwidth draws back. From the practice of view, it is our goal to achieve broad bandwidth even sacrifices some pulse energy. If a beam tilt of ±0.5 mm can be afforded routinely, an FEL with pulse energy exceeding 200 μJ and bandwidth up to nearly 0.4% can be delivered to users at 0.1 nm wavelength.

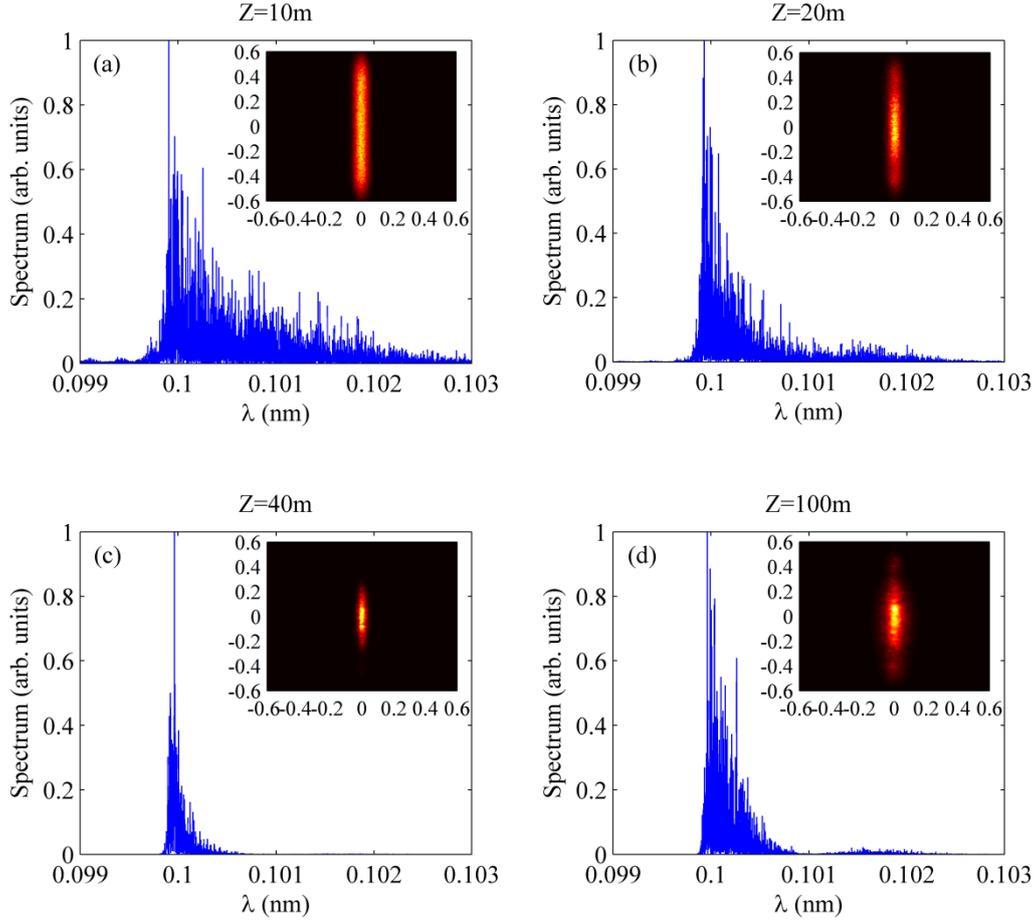

**Figure 6** Simulated spectrum and transverse profile of an XFEL pulse along the undulator: the four plots from (a) to (d) represents the four different locations of undulator.

With a vertical beam tilt of ±0.5mm, now we consider the radiation spectrum and spot evolution in the undulator, as presented in Fig. 6. Obviously, at the undulator position Z=10m, the generated spectrum in Fig. 6 (a) shows a bandwidth about 1%, and the radiation profile with a vertical envelope of 0.5 mm is consistent with the beam vertical tilt amplitude. Then as the electron beam continues to travel in the undulator, the radiation bandwidth gradually decreases. For instance, Fig. 6 (b) describes the radiation properties at the undulator position Z=20 m, i.e., in the exponential growth region. It can be found that the spectral bandwidth is narrowed and the radiation vertical size becomes shorter. Until the FEL lasing finally saturated which is presented in Fig. 6 (c), indeed, that FEL bandwidth sharply narrows to 0.25%, meanwhile, the vertical size of FEL spot goes to 0.2 mm which is much smaller than the initial situation. However after the saturation, as displayed in Fig. 6 (d), it is interesting to note that there is an increase of FEL bandwidth and enlarge of FEL vertical profile, accompanied with blurred periphery. At the exit of the undulator where Z=100 m, a FEL bandwidth of 0.4% and vertical FEL size approximately of 0.3 mm is obtained from simulation.

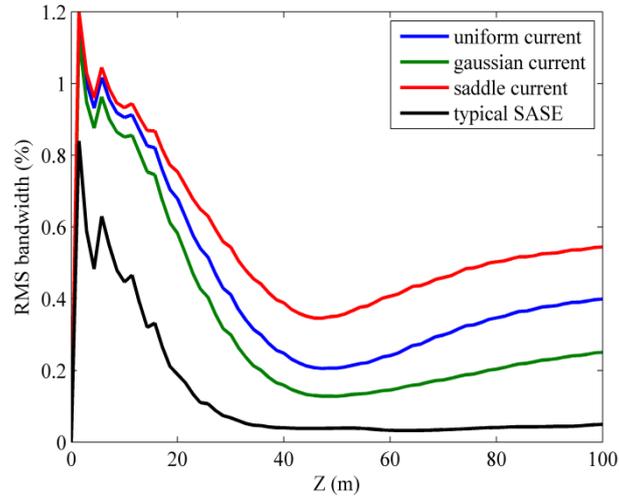

**Figure 7** FEL bandwidth evolutions along the undulator regarding to different bunch current distribution.

The blue curve of Fig. 7 displays the bandwidth evolution along the undulator under the vertical beam tilt amplitude equal to 0.5 mm, with the assumption of uniform or flat-top bunch current distribution. In principle, there exists several factors which contribute to the FEL bandwidth reduction along the undulator, e.g., the X-ray diffraction around the two edges which is usually larger than that of the centre are not compensated on time, and the electron exchange in the vertical plane cannot be avoided. More importantly, according to the Eq. (3), the FEL wavelength is determined by the square of beam vertical offset. Then if the electron beam has a vertical tilt amplitude of 0.5 mm, those electrons with the vertical position between 0.4 and 0.5 mm is responsible for producing 36% of the total bandwidth, while this part of beam only has 20% electrons under a flat-top current distribution assumption. Under such circumstance, it cannot afford a sufficient gain during the exponential growth when compared with those electrons in the centre, and leads a poor brightness of the 36% bandwidth. Therefore, there exists a sharp FEL bandwidth reduction before the initial saturation. After that, with the continuous gain of those unsaturated wavelength, the bandwidth experiences a slightly gradual growth. From the analysis, it is easy to figure out that an effective method to increase FEL bandwidth is using a saddle beam current distribution, and thus those insufficient FEL gain may be enhanced by the high current in the double horn. In modern FEL facilities, due to the bunch compress, a saddle bunch profile is a typical operation mode (Reiche *et al.*, 2005). For comparison, the Gaussian and saddle current distribution are taken into consideration and simulated. According to the plots in Fig. 7, it is definitely demonstrated that the saddle current beam achieve a bandwidth of 0.55%, while the Gaussian beam can only offer a bandwidth of 0.25%. However, when compared with typical SASE bandwidth of 0.05 % in the hard X-ray region (see the black curve in Fig. 7), it is still improved to an appreciable level and could be applied in a wide range of scientific researches.

## 4. Conclusion

Typically, the FEL sources aim to pursue the narrow bandwidth radiation. However, there is a strong demand to generate large bandwidth X-ray FEL pulse for several scientific applications. Therefore, in this paper, under the optimized undulator lattice, a transversely tilted beam is sent through the planar undulator with natural gradient to achieve the broad bandwidth FEL pulse. The electron beam vertical tilt can be obtained by the transverse wakefields of corrugated structure or a deflecting cavity, and then the FEL bandwidth is expected to be simply controlled by variation of the corrugated structure gap or the deflecting voltage. It is found that, during the FEL process, the produced radiation bandwidth sharply drops during the exponential growth, which is mainly caused by the unfair gains for different colour radiations. The numerical simulation demonstrated that, an electron beam with ±0.5 mm vertical tilt can deliver 0.1 nm FEL pulse with RMS bandwidth 0.25% to 0.55%, corresponding to different beam current profiles, which confirms the validity of the scheme and theoretical analysis used in this paper. More importantly, the novel scheme proposed here relies on the mature accelerator technique and the existing devices in modern short-wavelength FEL facilities. It paves the way for experiments requiring large bandwidth or multi-colour X-ray FEL pulse.